# A Chunk Caching Location and Searching Scheme in Content Centric Networking

Yang Li, Tao Lin, Hui Tang
High Performance Network Laboratory
Institute of Acoustics, Chinese Academy of Sciences
Beijing, China
{liy, lint, tangh}@hpnl.ac.cn

Peng Sun
School of Control Science and Engineering
Shandong University
Jinan, China
sunpengsdu@gmail.com

*Abstract*—Content Centric Networking (CCN) is a new network infrastructure around content dissemination and retrieval, shift from host addresses to named data. Each CCN router has a cache to store the chunks passed by it. Therefore the caching strategy about chunk placement can greatly affect the whole CCN performance. This paper proposes an implicit coordinate chunk caching location and searching scheme (CLS) in CCN hierarchical infrastructure. In CLS, there is at most one copy of a chunk cached on the path between a server and a leaf router. This copy is pulled down one level towards the leaf router by a request or pushed up one level towards the server by the cache eviction. Thus, it is possible to store more diverse contents in the whole CCN and improve the network performance. Plus, in order to reduce the server workload and file download time, a caching trail of chunk is created to direct the following request where to find the chunk. Extensive test-bed experiments have been performed to evaluate the proposed scheme in terms of a wide range of performance metrics. The results show that the proposed scheme outperforms existing algorithms.

## I. INTRODUCTION

With the spread of the information-centric services, the need for a content-aware infrastructure has been addressed to a certain extent through various solutions like CDN, P2P and HTTP proxies, deployed on top of the current Internet. In parallel, significant research projects have been funded in the recent years focusing on the definition of novel infrastructure for the future Internet (e.g. US NSF GENI). It is generally accepted that named data, instead of its physical location, is the central element of routing in the future Internet. Content Centric Networking (CCN) [1], one of the predominant proposals, radically changed data transfer by pushing content storage and delivery through name at network layer itself.

Chunk-level caching is a distinctive feature of CCN infrastructure and plays a fundamental role on system performance. Recently, there are two chunk-level caching analytical models proposed to evaluate data transfer in CCN using markov chains [2] and markov modulated rate process [3]. Furthermore, an analytical model of bandwidth and storage sharing is provided in [4] to guide a tradeoff between user performance and limited network resources. In addition, some experimental evaluations of storage management in CCN are presented in [5]. All the above related works, however, do not explore any optimal caching management strategies. Actually, dynamically placing the chunks in suitable intermediate nodes is an important and challenging task.

The optimal object placement algorithm needs the caches' coordination along the download path to the clients, which can be divided into two main types: explicit and implicit. With explicit coordination, caches share their state and additional information with each other [6, 7]. Using this information, each cache determines what to cache, when to do so and what to drop. In [6], the object placement problem is formulated as an optimization problem and the optimal locations to cache the object are obtained using additional information such as access patterns and content popularity. In addition, a *Filter* algorithm presented in [7], which aims at finding fundamental design principles for hierarchical caches, makes local caching decisions based on the document request frequency at the leaf cache which received the original request. However, the main cost of these explicit schemes is the additional communication overhead needed for coordination as well as coordination algorithms that can be quite complex and sophisticated.

Implicit coordination, on the other hand, removes the need for such elaborate reporting protocols. Instead, it relies on local cache management policies, as well as the relative position of each cache in the network, to achieve good performance [8-11]. Therefore, implicit coordination is more applicable than the explicit coordination. This is the reason why we focus on the implicit coordination scheme in this paper. Currently, it is not optimal to cache the download chunk at all intermediate routers in CCN. The implicit schemes in [10, 11], however, still have some problems. For example, placing the copy of a chunk at few routers may cause the other clients long download time, which will be discussed in details in the next section.

This paper proposes an implicit coordinate chunk caching location and searching scheme (CLS) in CCN hierarchical infrastructure. The hit chunk at level $l$ is pushed down to the $l$-1 level cache towards the clients, and the evicted chunk at level $l$ is pushed back to the $l$+1 level cache. Thus, there is always one and at most one copy of a chunk cached on the whole path between a server and a leaf router. Furthermore, a trail storing the chunk caching history is set up during the chunk cached up and down, which is used to definitely direct the following chunk search.

The main contributions of this paper are:

- We propose an implicit coordinate chunk placement scheme to improve the CCN cache efficiency.
- We provide a chunk searching policy based on the caching trail. At the best of our knowledge, this is the first scheme proposed by considering both caching strategy and request routing together.
- We evaluate the effectiveness of the CLS scheme by extensive test-bed experiments. The results show that our scheme outperforms existing schemes.

The remainder of this paper is organized as follows: Section II presents the background and relevant problems analysis, and then the proposed CLS scheme is described in detail in Section III. Scheme evaluation through test-bed results are presented in Section IV. Finally, Section V offers some concluding remarks.

## II. BACKGROUND

This section covers the introduction of CCN，three implicit coordinate caching schemes, and relevant problems analysis.

### A. CCN

It is generally recognized that the CCN has the following four main characteristics [4]: (1) named objects segmented in uniquely identified chunks; (2) receiver-driven chunk-based transport protocol; (3) routers with in-network per-chunk storage capabilities; (4) name-based routing and forwarding primitives. Here we give a simple CCN description as follows.

There are two CCN packet types, Interest and Data. Data is transmitted only in response to an Interest. The CCN router has three main data structures: the Forwarding Information Base (FIB), Content Store (CS) and Pending Interest Table (PIT). The FIB which is created based on the CCN name routing protocol, is used to forward Interest packets toward potential source(s). The CS caches the chunks passed by it according to the cache strategy. The PIT keeps track of Interests forwarded upstream toward content source(s) so that returned Data can be sent downstream to its original requester(s). The caching policy in the CCN is called *whole course caching*, which means a copy of the requested chunk is cached in all intermediate routers between hit point and requesting client.

### B. Breadcrumbs

An implicit, transparent and best-effort approach towards caching, called *Breadcrumbs*, is proposed in [9] to provide a simple content caching and searching. The *whole course caching* is also used in *Breadcrumbs*, like CCN. A trail for the purpose of storing routing history is created and maintained indefinitely at each router as the file is downloaded. Thus, the following request for the same file may be routed downstream towards the clients directing by such trail, instead of upstream towards the sources. Accordingly, a time thresholds is used in this scheme to judge whether searching downstream or not. Such downstream request routing approach may reduce the server workload and file download time if the file is found downstream. However, the request may also suffer missing downstream if the file is evicted at each router.

### C. LCD and MCD

A simple implicit coordinate caching scheme is called *Leave Copy Down* (LCD) [10]: under LCD a new copy of the requested document is cached only at the *l*-1 level cache, i.e., the one that resides immediately below the location of the hit on the path to the requesting client. Another scheme is called *Move Copy Down* (MCD) [10]: similar to LCD with the difference that a hit at level *l* moves the requested document to the underlay cache. The operation of the above mentioned schemes are illustrated in Fig. 1.

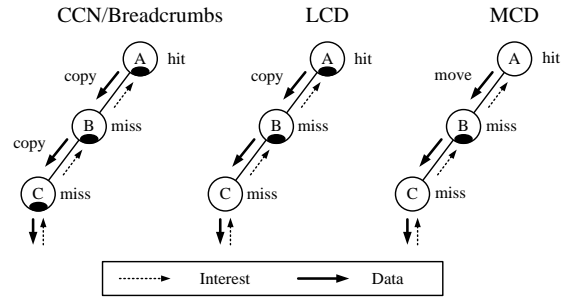

Figure 1. The cache operation of CCN, LCD and MCD.

According to the simulation results in [11], LCD performed better than MCD under the tree topologies in fact that LCD has one more copy to serve other clients connected to other branches. Furthermore, the simulation results also showed the LCD was superior to the CCN and the *Filter* algorithm.

### D. Problem Analysis

Comparing to *whole course caching*, LCD and MCD have the merits of being able to avoid the amplification of *replacement errors* and achieve the *cache exclusivity* [10]. *Replacement errors* occur when a less popular object causes the eviction of a more popular one. Leaving a copy in all the intermediate caches is, in effect, leading to the amplification of replacement errors. *Cache exclusivity* is achieved though reducing the unnecessary repetitious caching of the same objects at multiple levels.

On the other hand, *whole course caching*, like CCN, places copies at all the intermediate routers to achieve the following two goals: (1) have a nearby copy to service other clients connected to other branches that do not have a copy of the document; (2) have a "backup" copy for the requesting client in case its leaf copy is evicted from the leaf cache. LCD and MCD cannot achieve the two goals.

The request routing policy in *Breadcrumbs* is able to resolve the above contradiction to a certain degree. However, the *Breadcrumbs* still suffers a miss risk during searching downstream. In addition, the *Breadcrumbs* only focus on the searching algorithm without considering the chunk placement strategy. Therefore, this paper proposes an implicit coordinate CLS scheme to address the above contradiction by sharing some resemblance with MCD and *Breadcrumbs*.

## III. PROPOSED SCHEME

This section describes the proposed CLS scheme in details under CCN hierarchical infrastructure.

## A. Main Operation

The key idea of CLS is that a hit at level $l$ pulls the requested chunk down to the under level CS. And an eviction at level $l$ pushes the banished chunk back to the upper level CS. Meanwhile, a caching trail is created along the download path to assist chunk search. The details are as follows:

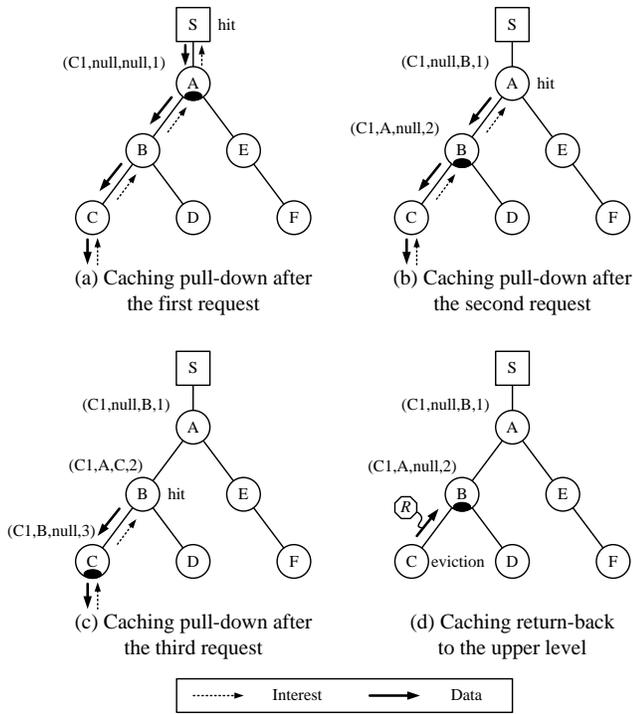

Figure 2. An example of the CLS scheme operation.

*1) Caching pull-down to the under level:* under CLS the hit chunk at level $l$ is pulled down to the $l$-1 level CS, the same operation as MCD. This requires that the requested chunk be deleted from the CS (moved to the bottom of the stack) where the hit occurred. No deletion of course takes place when the hit occurs at the server. CLS requires multiple requests to bring a chunk to a leaf router, with each request advancing a new copy of the chunk one hop closer to the client, as shown in Fig. 2 (a)-(c). Thus, there is at most only one copy of the chunk is cached on the path between a server and a leaf router.

The idea behind *Caching pull-down* is to avoid the amplification of *replacement errors* and achieve the *cache exclusivity* as mentioned above.

*2) Caching return-back to the upper level:* when a chunk is evicted from a CS at level $l$ due to replacement, for example LRU, the CLS scheme forces its caching at level $l$+1 with a Flag $R$, as shown in Fig. 2 (d). The Flag $R$ will be explained in next subsection. Accordingly, a least requested chunk may be pushed back up and up, till to the server. Except returning back to the server, the CLS ensures that there is always one copy of the chunk cached on the path between a server and a leaf router.

Thus, the assurance of existence of such one copy is able to direct the following Interest right to the cached router, instead of the server, consequently leading to a reduction on server workload and download time. The searching policy is explained as follows.

*3) Searching policy according to the trail:* A caching trail is created at each router along the download path in CLS. Each trail is a 4-tuple entry (ID, *in*, *out*, *h*), indexed by a global unique chunk ID, containing the following information: (1) Face of incoming router from which the chunk arrived, *in*; (2) Face(s) of outgoing router(s) to which the chunk was pushed down, *out*; (3) The number of hops from the server to this router, *h*. Each router modifies the PIT for the purpose of storing the trail information of previously cached chunks.

The key difference of trails between CLS and *Breadcrumbs* is that the CLS trail is created at the time the chunk is cached, not passed by. An example of the trail creating procedure is illustrated in Fig. 2. When the first Interest pulls the chunk to A from server, a trail is created at A as (C1, null, null, 1), as shown in Fig. 2 (a). The *null* value of *in* means that the chunk is arrived from server, and the *null* value of *out* means that the chunk is cached locally without being pushed down. The value of *h*, carried in the Data chunk, increases one after advancing one hop. Then, with the second Interest, the chunk is pushed down to B. So the trail at A is modified to (C1, null, B, 1), while the trail at B is created as (C1, A, null, 2) (see Fig. 2 (b)). Accordingly, the third Interest changes the trails at A, B and C as shown in Fig. 2 (c). On the other hand, the trail will be deleted if the chunk is evicted by replacement (see Fig. 2 (d)). In such a case, the trail at B is also modified to (C1, A, null, 2) after caching the evicted chunk with Flag $R$.

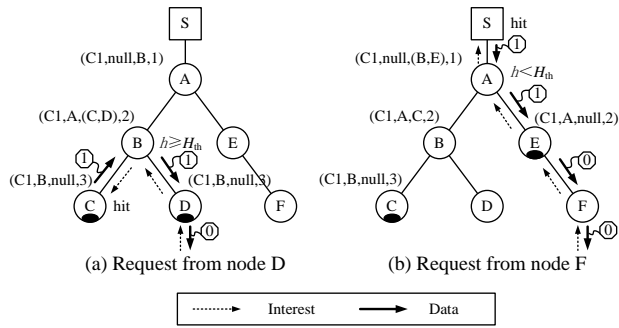

Figure 3. An example of the searching policy according to the trail ($H_{th}$ =2).

Therefore, the existence of a trail at a router means that there is a copy of such chunk cached on it or downstream, which is able to be used to direct the following Interest. An obvious question is in which direction to search a chunk: upstream to the server or downstream to the cached copy. Here, a simple policy is provided by comparing the number of the hops ($h$) with a predefined threshold ($H_{th}$). It is reasonable to set the value of $H_{th}$ as the half of the number of hops from the server to the leaf router according to the ISP practical situation, which is out of the scope of this paper. Then, an intermediate router would forward an Interest downstream according to trails, instead of upstream according to FIB, if-and-only-if $h$ is not less than the $H_{th}$. An example of this searching policy is showed in Fig. 3 where $H_{th}$ is defined to be 2. Consequently, in order to search the same chunk (C1), the router B forwards the Interest from D downstream (see Fig. 3 (a)), while the router A forwards the Interest from F upstream (see Fig. 3 (b)).

It is obvious that the CLS is superior to *Breadcrumbs* in that the CLS can ensure successful finding a chunk if search downstream, and save cache space by deleting a trail on time.

### B. Special Cases

In addition, there are some special cases should be explained in order to ensure the CLS works correctly. To achieve the cache exclusive, it requires that the CLS cache only one copy of one chunk on the path between a server and a leaf router. This is the design principle for following special cases.

Before describing the details, let's first illustrate the Flag *R*. In CLS, each Data is added a Flag *R*, which is used to identify the chunk state. More precisely, the value of *R* is set to 1 when a chunk is hit and modified to 0 after it is cached, as shown in Fig. 3. When a chunk is evicted, the value of *R* is set to 2.

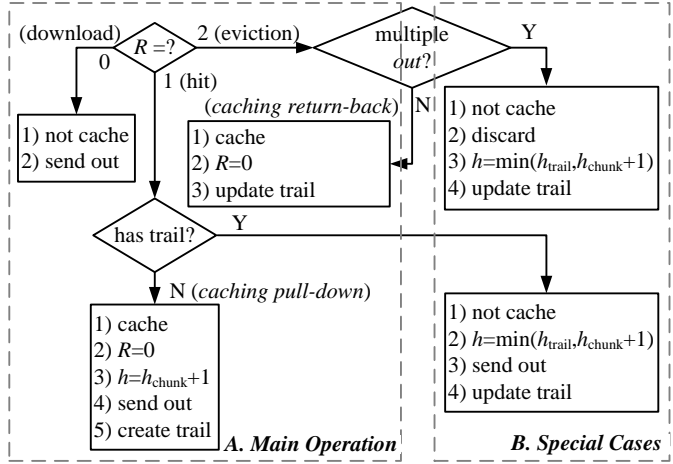

Figure 4. The operation of a router in CLS scheme when receiving a Data.

One special case occurs when the router who has a trail receives a chunk with *R*=1, which differs from the above *caching pull-down* in that the chunk is sent out immediately without caching, as shown in bottom right of the Fig. 4. An example of such case is illustrated in Fig. 3 (a), where the router B sends the hit chunk to the D without caching, ensuring only one copy of the chunk on path S-A-B-C and path S-A-B-D. Accordingly, the router B should update their trails by adding the D in the *out*. Plus, the value of *h* should be set to the smaller one between that in the trail and that derived from the received chunk ($h=\min(h_{trail}, h_{chunk}+1)$). As shown in this example, when receiving a chunk with $h_{chunk}$=3 from C, the router B, who has a trail with $h_{trail}$=2, sets the value of *h*=2 to its trail and the chunk. Furthermore, it should be pointed out that a hit chunk is not deleted from CS if it is sent out to the previous incoming face *in* according to the trail, like C in Fig. 3 (a).

Another special case occurs when an evicted chunk is pushed back to a router who has a trail with multiple *out* faces. The operation in this case is different from the above *caching return-back* in that the receiver just discards the chunk without caching, as shown in top right of the Fig. 4. An example of this case is shown in Fig. 5, where the router B deletes the evicted chunk (*R*=2) since its *out* value indicates that there is one copy of the chunk cached on one path passed B (S-A-B-D). As a result, the router B updates its trail by deleting C from *out* faces.

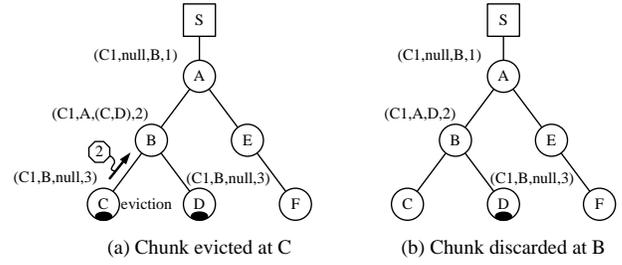

Figure 5. The router B operation upon receiving evited chunk when B has two outgoing routers.

## IV. SCHEME EVALUATION

### A. Experimental Results

Extensive test-bed experiments have been performed to compare the CLS with existing schemes, CCN and LCD. The experimental evaluation is performed on our modified version of CCNx prototype [13], where the topology is the same as that shown in Fig. 2. We consider 100 files, equally partitioned into 100 classes of average size 50 KB, and the chunk size is 4kBytes. In order to build fairly realistic network condition, the content popularity distribution is assumed to be Zipf($\alpha$) with $\alpha$= 0.3 or 0.9. Values of $\alpha$ close to 1 indicate that few distinct contents attract the majority of the requests while values close to 0 indicate almost uniform document popularities [13]. Users are connected to leaf router and generate the content request according to a Poisson process of intensity $\lambda$= 5 req/s [14]. All routers are equipped with an LRU CS, the total size of these CSs ranged from 10% to 50% of the whole contents.

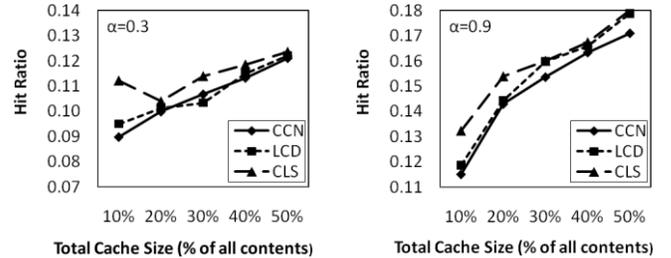

Figure 6. Hit ratio vs. cache size.

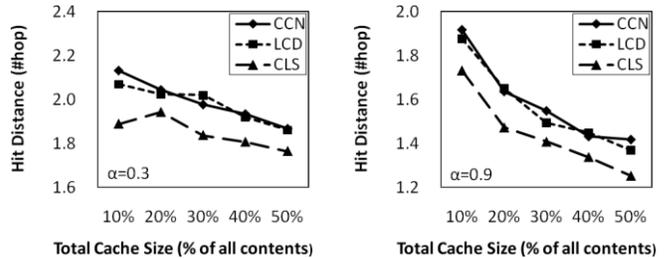

Figure 7. Hit distance vs. cache size.

Fig. 6 plots the average hit ratio of all the routers curves as a function of cache size for different caching schemes. The hit ratio is defined as the ratio of the number of Interest served by the caches to the total number of Interests arrived at the cache. The CLS scheme improves the chunk hit ratio over the other schemes. For example, at cache size 20% ($\alpha$= 0.9), the hit ratios are 14.28%, 14.43%, 15.4% for CCN, LCD, CLS respectively. This also implies substantial load reduction at the content servers. Fig. 7 shows the average hit distance under each scheme, where the hit distance is measured in number hops from the client to the hit cache. The lower the average hit distance, the better the performance. The CLS scheme reduces the hit distance (about 12%) compared to the other schemes. This is not surprising because the CLS can reduce the *replacement error* and achieves *cache exclusivity* through keeping only one copy of the chunk cached on the path.

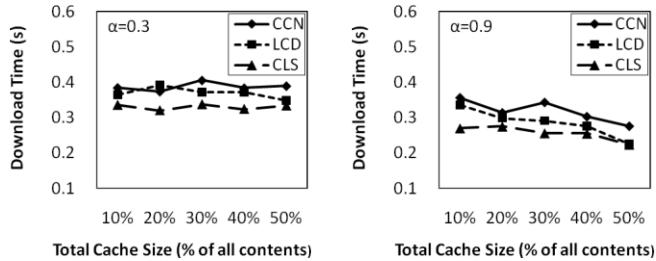

Figure 8. Download time vs. cache size.

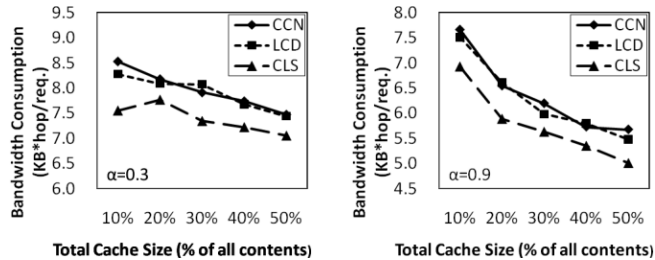

Figure 9. Bandwidth consumption vs. cache size.

Fig. 8 depicts the download time. By using the searching policy proposed in this paper, the CLS scheme slightly reduces the download time over the other schemes. By contrast, the CCN and LCD introduce about 50 ms more than that in CLS because some chunks may be found downstream in CLS. Finally, we examine the load on the network links. Fig. 9 shows the average network traffic (measured in byte*hops) required to satisfy a request. From these figures, it can be clearly seen that the CLS scheme results in much lower load on the network links than the other schemes.

In addition, the results reveals that the CLS only provides a little better performance than the LCD and CCN. The reason is that all results are derived from a three-level test-bed, which has a limitation to reflect the superiority of the CLS schemes. Furthermore, it can be clearly seen from all the above experimental results that the scenario of α= 0.9 show a better performance than that of α= 0.3 for all three schemes. This is because the benefits of such cache-and-forward architectures are more obvious when majority clients require few contents.

### B. Discussion

It should be pointed that the CLS achieves the above merits at a cost of maintaining trails and keeping the faces information of the neighboring caches at each router. In addition, the CLS still has some limitations which are the focus of our future work. For instance, the above searching policy in the CLS cannot ensure the chunk be found at the nearest cache point to the client sometimes. The reason is that an intermediate router cannot judge who is close, the cached copy or server, according to the trail. Moreover, it is worth validating the CLS performance in a large architecture with many levels, such as 10 levels. It may be more preferable to place more than one copy with large interval on the download path in that case.

### V. CONCLUSION

This paper proposed an implicit coordinate chunk placement strategy and a simple Interest searching policy in CCN hierarchical infrastructure. Through ensuring only one copy of one chunk on a path, the CLS scheme could reduce replacement errors and achieve the cache exclusivity. With a simple searching policy, the CLS scheme could reduce the server workload and file download time. The experimental results showed that the CLS scheme effectively increases hit ratio and reduces hit distance, download time and bandwidth consumption compared to the existing caching algorithms.

ACKNOWLEDGMENT

This work is supported by the National Science & Technology Major Project of China (No. 2011ZX03002-001-01) -- The research on the mobile Internet architecture, and the National Nature Science Foundation of China (No. 61100178 and No. 11161140319).